\documentclass[conference]{IEEEtran} 
\usepackage{hyperref} 
\usepackage{graphicx}
\usepackage{amssymb}
\usepackage{amsmath}
\usepackage{mathtools}
\usepackage{cite}
\usepackage{stfloats}
\usepackage{subfigure}
\usepackage{psfrag}
\usepackage[mathscr]{euscript}
\usepackage{acronym}  
\usepackage{booktabs} 
\usepackage[table]{xcolor}
\usepackage{slashbox,pict2e}
\usepackage{dsfont}

\acrodef{CCDF}{complementary cumulative distribution function}
\acrodef{CF}{characteristic function}
\acrodef{PPP}{Poisson point process}
\acrodef{CSI}{channel state information}
\acrodef{OFDM}{orthogonal frequency division multiplexing}
\acrodef{OFDMA}{orthogonal frequency division multiple access}

\acrodef{RV}{random variable}
\acrodef{i.i.d.}{independent, identically distributed}
\acrodef{PMF}{probability mass function}
\acrodef{PDF}{probability distribution function}
\acrodef{CDF}{cumulative distribution function}
\acrodef{ch.f.}{characteristic function}
\acrodef{AWGN}{additive white Gaussian noise}
\acrodef{SNR}{signal-to-noise ratio}
\acrodef{LRT}{likelihood ratio test}
\acrodef{DRT}{distance ratio test}
\acrodef{GLRT}{generalized likelihood ratio test}
\acrodef{CRLB}{Cram\'{e}r-Rao lower bound}
\acrodef{CRB}{Cram\'{e}r-Rao bound}
\acrodef{ZZLB}{Ziv-Zakai lower bound}
\acrodef{ZZB}{Ziv-Zakai bound}
\acrodef{LoS}{line-of-sight}
\acrodef{ToF}{time-of-flight}
\acrodef{NLoS}{non-line-of-sight}
\acrodef{GDOP}{geometric dilution of precision}
\acrodef{GPS}{Global Positioning System}
\acrodef{FIM}{Fisher information matrix}
\acrodef{PEB}{position error bound}
\acrodef{SPEB}{squared position error bound}
\acrodef{TOA}{time-of-arrival}
\acrodef{TOF}{time-of-flight}
\acrodef{WSN}{wireless sensor network}
\acrodef{MAC}{medium access control}
\acrodef{RSS}{received signal strength}
\acrodef{WAF}{wall attenuation factor}
\acrodef{TDOA}{time difference-of-arrival}
\acrodef{RF}{radiofrequency}
\acrodef{RTT}{round-trip time}
\acrodef{AOA}{angle-of-arrival}
\acrodef{MF}{matched filter}
\acrodef{ED}{energy detector}
\acrodef{ML}{maximum likelihood}
\acrodef{MSE}{mean-square error}
\acrodef{RMSE}{root-mean-square error}
\acrodef{LEO}{localization error outage}
\acrodef{ppm}{part-per-million}
\acrodef{ACK}{acknowledge}
\acrodef{UWB}{Ultrawide bandwidth}
\acrodef{TNR}{threshold-to-noise ratio}
\acrodef{LS}{least squares}
\acrodef{IR-UWB}{impulse radio UWB}
\acrodef{FCC}{Federal Communications Commission}
\acrodef{TH}{time-hopping}
\acrodef{PPM}{pulse position modulation}
\acrodef{MUI}{multi-user interference}
\acrodef{PDP}{power delay profile}
\acrodef{BPZF}{band-pass zonal filter}
\acrodef{SIR}{signal-to-interference ratio}
\acrodef{RFID}{radio frequency identification}
\acrodef{WPAN}{wireless personal area network}
\acrodef{WWB}{Weiss-Weinstein bound}
\acrodef{DP}{direct path}
\acrodef{MF}{matched filter}
\acrodef{MMSE}{minimum-mean-square-error}
\acrodef{SBS}{serial backward search}
\acrodef{SBSMC}{serial backward search for multiple clusters}
\acrodef{NBI}{narrowband interference}
\acrodef{WBI}{wideband interference}
\acrodef{INR}{interference-to-noise ratio}
\acrodef{CR}{channel response}
\acrodef{CIR}{channel impulse response}
\acrodef{CR}{channel  response}
\acrodef{RADAR}{radar}
\acrodef{MUR}{Multistatic radar}
\acrodef{JBSF}{jump back and search forward}
\acrodef{HDSA}{high-definition situation-aware}
\acrodef{RRC}{root raised cosine}
\acrodef{ST}{simple thresholding}
\acrodef{BTB}{Bellini-Tartara bound}
\acrodef{P-Max}{$P$-Max}  
\acrodef{MIMO}{multiple-input multiple-output}
\acrodef{MAP}{maximum a posteriori}
\acrodef{FG}{factor graph}
\acrodef{OP}{outage probability}
\acrodef{WED}{wall extra delay}
\acrodef{RMS}{root mean square}
\acrodef{SPAWN}{sum-product algorithm over a wireless network}
\acrodef{MDD}{minimum distance distribution}
\acrodef{MAP}{maximum a posteriori probability}
\acrodef{PAR}{probabilistic association rule}
\acrodef{AP}{access point}
\acrodef{HD}{half-duplex}
\acrodef{FD}{full-duplex}
\acrodef{IC}{interference cancellation}
\acrodef{HDHN}{hybrid-duplex heterogeneous network}
\acrodef{TDD}{time-division duplexing}
\acrodef{FDD}{frequency-division duplexing}
\acrodef{SINR}{signal-to-interference-plus-noise ratio}
\acrodef{UAV}{unmanned aerial vehicle}
\acrodef{GCS}{ground control station}
\acrodef{LTE}{long term evolution}
\usepackage{color}
\usepackage{dsfont}
\usepackage{bbm}

\newcommand{\PL}[1]{\alpha_{#1}} 

\newcommand{\Pm}{P_\text{m}}
\newcommand{\PI}{P_\text{I}}

\newcommand{\Lm}{\ell_\text{m}}
\newcommand{\Lma}{\ell_\text{m}^{-\alpha_\text{m}(\theta_\text{m})}}
\newcommand{\Li}{\ell_\text{I}}
\newcommand{\Lia}{\ell_\text{I}^{-\alpha_\text{I}(\theta_\text{I})}}
\newcommand{\Hm}{h_\text{m}}
\newcommand{\Hi}{h_\text{I}}
\newcommand{\HL}{\overline{H_\text{L}}}
\newcommand{\HN}{\overline{H_\text{N}}}

\newcommand{\Km}{K_\text{m}}
\newcommand{\Ki}{K_\text{I}}

\newcommand{\Ws}[2]{{W_{}^{}}} 

\newcommand{\TSIR}[2]{{\tau_{}^{}}}

\newcommand{\Pout}{p_{\text{o}}}

\newcommand{\PLoS}{p_{\text{L}}(\theta_i)}
\newcommand{\PLoSm}{p_{\text{L}}(\theta_\text{m})}
\newcommand{\PLoSi}{p_{\text{L}}(\theta_\text{I})}

\DeclareMathAlphabet{\mathsf}{OML}{cmbr}{m}{it}

\newtheorem{theorem}{Theorem}

\newtheorem{corollary}{Corollary}

\newcommand{\bd}{\begin{description}}
	\newcommand{\ed}{\end{description}}
\newcommand{\be}{\begin{enumerate}}
	\newcommand{\ee}{\end{enumerate}}
\newcommand{\bi}{\begin{itemize}}
	\newcommand{\ei}{\end{itemize}}
\newcommand{\bl}{\begin{list}}
	\newcommand{\el}{\end{list}}
\newcommand{\bt}{\begin{tabbing}}
	\newcommand{\et}{\end{tabbing}}

\newcounter{eqncnt}

\newcounter{eqnback}

\acrodef{BS}{base station}
\acrodef{G2A}{ground-to-air}
\acrodef{A2G}{air-to-ground}
\acrodef{A2A}{air-to-air}
\acrodef{G2G}{ground-to-ground}
\acrodef{IoT}{internet of things}

\IEEEoverridecommandlockouts 

\begin{document}

\newcommand{\paperTitle}{Outage Probability of UAV Communications\\ in the Presence of Interference}
%
 
 

\title{\paperTitle}


\author{
	\IEEEauthorblockN{
		Minsu~Kim and 
		Jemin~Lee
	}\\[-0.4em]
	\IEEEauthorblockA{
		Department of Information and Communication Engineering (ICE)\\
		Daegu Gyeongbuk Institute of Science and Technology (DGIST), Korea\\
		Email: {ads5577@dgist.ac.kr}, {jmnlee@dgist.ac.kr}
	}
	\thanks{
		This work was supported in part by the the National Research Foundation of Korea (NRF) grant funded by the Korea government (MSIP) (No. 2017R1C1B2009280) and the DGIST R\&D Program of the Ministry of Science and ICT(17-ST-02).
	}
}
\maketitle 
%

%

%
\setcounter{page}{1}
\acresetall
\begin{abstract}
Unlike terrestrial communications, \ac{UAV} communications have some advantages such as \ac{LoS} environment and flexible mobility. However, the interference will be still inevitable. 
In this paper, 
we analyze the effect of the interference on the \ac{UAV} communications
by considering
the \ac{LoS} probability and different channel fadings for \ac{LoS} and \ac{NLoS} links, 
which are affected by the elevation angle of the communication link.
	%
	%
We then derive a closed-form outage probability in the presence of an interfering node for all the possible scenarios and environments of main and interference links.
After discussing the impacts of transmitting and interfering node parameters on the outage probability, 
we show the existence of the optimal height of the \ac{UAV} that minimizes the outage probability. 
We also show the \ac{NLoS} environment can be better than the \ac{LoS} environment if the average received power of the interference is more dominant than that of the transmitting signal in \ac{UAV} communications.
\\[-0.5em]
\end{abstract}
%
\acresetall
%
%
\section{Introduction}\label{sec:Intro}
%
As the \ac{UAV} technology develops, reliable \ac{UAV} communications have become necessary.
However, since \ac{UAV} communications are different from conventional terrestrial communications, it is hard to apply the technology used in terrestrial communications to \ac{UAV} communications\cite{ZenZhaLim:16}.
Especially, unlike terrestrial communications, \ac{UAV} communications can have \ac{LoS} environments between a \ac{UAV} and a ground device, and between \acp{UAV}.
When the main link is in a \ac{LoS} environment, the received main signal power will increase due to better channel fading and lower path loss exponent compared to a \ac{NLoS} environment. 
It also means that in the presence of an interfering node, the interfering signal can be received with lager power as the interfering link can also be in a \ac{LoS} environment \cite{ChoLiuLee:18}.

%
%
\ac{UAV} communications have been studied in the literature, mostly focused on the optimal positioning and trajectory of the \ac{UAV}.
The height of the \ac{UAV} affects the communication performance in different ways.
As the height increases, the \ac{UAV} forms the \ac{LoS} link with higher probability, which is modeled by the \ac{LoS} probability in \cite{ITU:13}, but the distance to the receiver at the ground increases as well.
By considering this relation, the optimal height of the \ac{UAV} in terms of the communication coverage in the \ac{A2G} channel is presented in \cite{HouKanLar:14},
and for the case of using an \ac{UAV} as a relay, the optimal height and position of \acp{UAV} have also benn presented in \cite{AzaRoChePol:18}.
The work \cite{ZhaZhaHe:18} jointly optimized \ac{UAV} trajectory and power control to minimize the outage probability without considering the \ac{LoS} probability.
However, all of those works analyzed and optimized for the \ac{UAV} communications in the absence of an interfering node. Since the interference is an inevitable factor in the current and future networks, the impact of the interference on the \ac{UAV} communications needs to be investigated carefully.

Recently, the interference has been considered in some works for the optimal positioning and trajectory of the \ac{UAV}.
The optimal deployment of the \ac{UAV} has been presented to maximize the communication coverage in \cite{CheDhi:17,MozSaaBen:16}.
The user scheduling and \ac{UAV} trajectory have been jointly optimized with maximizing the minimum average rate without considering the \ac{LoS} probability in \cite{WuZenZha:18},
and the \ac{UAV} trajectory is also optimized jointly with device-\ac{UAV} association and uplink power to minimize the total transmit power according to the number of update times in \cite{MozSaaBenDeb:17}.
%
However, all of those prior works considered limited \ac{UAV} communication scenarios or environments.
Specifically, only the path loss is used for channels without fading in \cite{MozSaaBen:16, WuZenZha:18, MozSaaBenDeb:17},
or the fact that the \ac{LoS} probability can be different according to the locations of the \ac{UAV} was not considered in \cite{CheDhi:17}.
%

%
%
%
Therefore, in this paper, we analyze the effect of the interference on the \ac{UAV} communications by considering both the \ac{LoS} and \ac{NLoS} links and channel fading.
The probability of forming the \ac{LoS} link is defined by the elevation angle between a \ac{UAV} and a ground device, and the path loss exponent and the Rician factor are also determined differently by the elevation angle. 
The main contribution of this paper can be summarized as follows:
%
%
\begin{itemize}
	\item we consider all the scenarios of main (i.e., from a transmitter to a receiver) and interference (i.e., from an interfering node to a receiver) links in \ac{UAV} communications, which includes \ac{G2A}, \ac{G2G}, \ac{A2G}, and \ac{A2A} channels for the main and interference links;
	\item we derive a closed-form outage probability in the presence of interfering node for all the scenarios by considering the \ac{LoS} probability and different channel fading for \ac{LoS} and \ac{NLoS} links; and 
	%
	%
	\item we analyze how the heights of transmitting or interfering node and link distances affect the outage probability through numerical results.  
	%
\end{itemize}

%
%
%
%
%
\section{System Model}\label{sec:models}
In this section, we describe the network model and the channel model for \ac{UAV} communications.
\subsection{Terrestrial \& Aerial Network Models}

We consider a \ac{UAV} network, which has a \ac{UAV}, a ground device (e.g., ground control station or base station), and an interfering node. 
In this network, there can be three types of communications: \ac{UAV} to \ac{UAV}, 
\ac{UAV} to ground device (or ground device to \ac{UAV}), and ground device to ground device.
The interfering node can be either on the ground or in the air, and we consider one interfering node. \footnote{Note that when the multiple interfering nodes are considered, the communication performance such as the outage probability has the similar trend as only dominant interfering node is considered and it is generally determined by the dominant interfering node at the low outage region \cite{MorLoy:09}.}

\begin{figure}[t!]
	\begin{center}   
		{ 
			\psfrag{I}[Bl][Bl][0.59]{$\ell_I$}
			\psfrag{M}[Bl][Bl][0.59]{$\ell_m$}
			\includegraphics[width=1.00\columnwidth]{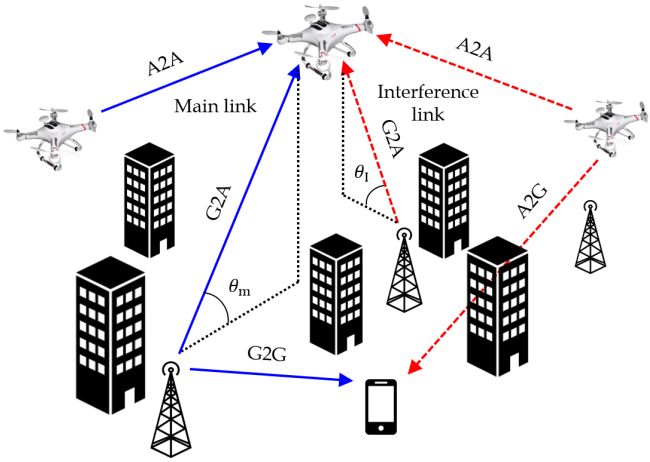}
			\vspace{-10mm}
		}
	\end{center}
	\caption{
		System model when UAVs are the communication devices. There are four types of channels: ground-to-ground (G2G), ground-to-air (G2A), air-to-ground (A2G), and air-to-air (A2A) channels. The blue lines represent the main links and the red dotted lines represent the interference links, and $\theta_\text{m}$ and $\theta_\text{I}$ are the elevation angle of main link and interference link, respectively.
	}
\vspace{-1mm}
	\label{fig:system}
\end{figure}
%
\par
When a transmitter (Tx), located at $(x_\text{m}, y_\text{m}, z_\text{m})$, communicates to a receiver (Rx), located at $(0,0, z_\text{o})$ in the presence of interfering node at $(x_\text{I}, y_\text{I}, z_\text{I})$, 
\ac{SIR} is given by
%
%
%
%
\setlength{\abovedisplayskip}{6pt}
\setlength{\belowdisplayskip}{6pt}
\begin{align} \label{eq:SIR}
	\gamma(\theta_\text{m},\theta_\text{I})&=
	\frac{\Hm \Lma \Pm}  {\Hi \Lia \PI} 
	= \frac{\Hm \beta_\text{m}(\theta_\text{m})}  {\Hi \beta_\text{I}(\theta_\text{I})} 
\end{align}
where $\beta_\text{m}(\theta_\text{m})$ and $\beta_\text{I}(\theta_\text{I})$ are respectively given by 
\begin{align}	
&\beta_\text{m}(\theta_\text{m})=\Lma \Pm, \quad
\beta_\text{I}(\theta_\text{I})=\Lia \PI.
\end{align}
Here, $\Hm$ and $\Hi$ are the fading gains of the main link (i.e., the channel between Tx and Rx) and the interference link (i.e., the channel between interfering node and Rx), respectively;
$\Lm=\sqrt{x_\text{m}^2+y_\text{m}^2+(z_\text{m}-z_\text{o})^2}$ and 
$\Li=\sqrt{x_\text{I}^2+y_\text{I}^2+(z_\text{I}-z_\text{o})^2}$ 
are the distances of main link and interference link, respectively; 
$\Pm$ and $\PI$ are the transmission power of the transmitter and the interfering node, respectively; and 
$\PL{\text{m}}(\theta_\text{m})$ and $\PL{\text{I}}(\theta_\text{I})$ are the path loss exponents of main link and interference link, respectively.
In \eqref{eq:SIR}, most parameters are determined by $\theta_\text{m}$ and $\theta_\text{I}$,
which are the elevation angles between Tx and Rx and between Rx and the interfering node, respectively, which are given by
\begin{align}	
\theta_i 
= 
\arctan 
\left( 
\frac{d_i^{(\text{V})}}{d_i^{(\text{H})}} 
\right), 
\quad 
\forall i = \{ \text{m}, \text{I}\}
\end{align}
where $d_i^{(\text{H})} = \sqrt{x_i^2 + y_i^2}$ is the horizontal distance
and $d_i^{(\text{V})} = \sqrt{(z_i - z_\text{o})^2}$ is the vertical distance of the main link ($i = \text{m}$) or the interference link ($i = \text{I}$).
\subsection{Channel Model} \label{subsec:channel}
As shown in Fig. \ref{fig:system}, there are three types of the channels in the UAV networks: 
the \emph{\ac{A2G} channel} (from \ac{UAV} to a ground device), the \emph{\ac{A2A} channel} (from \ac{UAV} to \ac{UAV}), and the \emph{\ac{G2G} channel} (from a ground device to a ground device).
The \ac{G2G} channel is the same channel of a terrestrial network, which is generally modeled as \ac{NLoS} environments with Rayleigh fading in urban area. 
The \ac{G2A} channel and the \ac{A2G} channel have the same characteristics.
Hence, we describe characteristics of the \ac{A2G} and \ac{A2A} channels in this subsection.


The \ac{A2G} and \ac{A2A} channels can have \ac{LoS} or \ac{NLoS} environments depending on the height of the \ac{UAV} and its surrounding environments such as buildings.
The elevation angle $\theta_i$ ($\theta_\text{m}$ or $\theta_\text{I}$) is considered for the \ac{A2G} (or \ac{G2A}) channel, while ignored for \ac{G2G} or \ac{A2A} channel and assumed to be $\theta_i = 0$ or $\frac{\pi}{2}$ for those two cases. In the following, we first describe the channel components, affected by $\theta_i$, and then provide the models for \ac{A2G} and \ac{A2A} channels.
\subsubsection{Components affected by elevation angle $\theta_i$} \label{subsec:component}
The elevation angle $\theta_i$ affects the probability of forming \ac{LoS}, the path loss exponent, and the Rician factor as described below.
%
%
%
\bi
\item 
The \emph{\ac{LoS} probability} is given by \cite{ITU:13}
\begin{align} \label{eq:LoS}
\PLoS=
\frac{1} {1 + a_1 \exp\left\{-b_1\left(\theta_i - a_1\right)\right\}}
\end{align}
where $a_1$ and $b_1$ are environment parameters, determined by the building density and height.
%
%
%
\item 
The \emph{path loss exponent} is determined by $\theta_i$ as \cite{AzaRoChePol:18}
\begin{align} \label{eq:pathloss}
\alpha(\theta_i)=
a_2 \PLoS + b_2
\end{align}
where 
$a_2=
\frac{\alpha({\frac{\pi} {2}})  -  \alpha(0)}
{p_\text{L}(\frac{\pi} {2})  -  p_\text{L}(0)}
\approx \alpha({\frac{\pi} {2}})  -  \alpha(0)$ and  
$b_2=
\alpha(0)  -  a_2 p_\text{L}(0)
\approx \alpha(0)$.
%
%
%
%
%
\item 
The \emph{Rician factor} is determined by $\theta_i$ as \cite{AzaRoChePol:18}
\begin{align}	\label{eq:k-factor}
K(\theta_i)
= a_3 \exp(b_3 \theta_i)
\end{align}
where 
$a_3=K(0)$ and
$b_3=
\frac{2}  {\pi}  
\ln\left(\frac{K({\frac{\pi}{2}})}  {K(0)}\right)$.
%
%
\ei
Note that from \eqref{eq:LoS}-\eqref{eq:k-factor}, we can see that $\PLoS$ and $K(\theta_i)$ are increasing functions of $\theta_i$ and $\alpha(\theta_i)$ is a decreasing function of $\theta_i$, so the received power increases when $\theta_i$ increases.
\subsubsection{Air-to-Ground (A2G) channel}
When the main link and the interference link are both \ac{A2G} channels, $\Hm$ and $\Hi$ can be in either \ac{LoS} or \ac{NLoS} environments.
We consider that the channel fading is Rician fading for \ac{LoS} environments and Rayleigh fading for \ac{NLoS} environments.
Therefore, the distribution of the channel fading, $h_i$, $i \in  \{\text{m}, \text{I}\}$, is given by
\begin{align}
f_{h_i}(h)=	
\left\{
\begin{aligned}
&
f_\text{L}(h) \quad\quad \text{for \ac{LoS} case} \label{eq:channel fading}\\
&
f_\text{N}(h) \quad\quad \text{for \ac{NLoS} case}
\end{aligned}
\right.
\end{align}
where $f_\text{L}(h)$ and $f_\text{N}(h)$ are noncentral Chi-squared and exponential distribution, respectively, and given by
\begin{align}	
f_\text{L}(h)
&=\frac{1 + K(\theta_i)}  {\HL}  
\exp\left(-K(\theta_i)-\frac{1+K(\theta_i)}  {\HL} h \right)\nonumber\\
&\quad\times I_0 \left(2\sqrt{\frac{K(\theta_i) (1 + K(\theta_i)) }  {\HL} h} \right)\nonumber\\
&=\frac{1}  {2}  
\exp\left(-K(\theta_i) - \frac{h}  {2}\right) 
I_0 \left(\sqrt{2 K(\theta_i) h} \right) \label{eq:channel fading1}\\
f_\text{N}(h)
&=\frac{1}  {\HN}
\exp\left(-\frac{h}  {\HN}\right)
=\exp\left(-h\right). \label{eq:channel fading2}
\end{align}
Here, $I_0(\cdot)$ is the modified Bessel function of the first kind with order zero,
and $\HL=2+2K(\theta_i)$ and $\HN=1$ are the means of \ac{LoS} and \ac{NLoS} channel fading gain, respectively.
\subsubsection{Air-to-Air (A2A) channel}
In \ac{A2A} channel, 
the channel will be in \ac{LoS} environments and $\theta_i = \frac{\pi}{2}$, so the distribution of the channel fading, 
$h_i$, $i \in  \{\text{m}, \text{I}\}$, is given by
\begin{align}
f_{h_i}(h)&=
\frac{1}  {2}  
\exp\left(-K_\text{o} - \frac{h}  {2}\right) 
I_0 \left(\sqrt{2 K_\text{o} h} \right)  \label{eq:noncent}
\end{align}
where $K_\text{o}=K(\frac{\pi}{2})$. Unlike the \ac{A2G} channel, 
the Rician factor $K_o$ and the path loss exponent $\alpha$ of \ac{A2A} channel are not affected by $\theta_i$ \cite{GodWie:15}.
%
%
\setcounter{eqnback}{\value{equation}}
\setcounter{equation}{12}
\begin{figure*}[t!]
	%
	%
	\begin{align}
	\Pout^{(\text{L,L})}(\Theta,\mathcal{D})
	&=  1  -  Q\left(\sqrt{\frac{2 \Km(\theta_\text{m}) \beta_\text{m}(\theta_\text{m})}  
		{\beta_\text{m}(\theta_\text{m}) + \gamma_\text{t} \beta_\text{I}(\theta_\text{I})}}  , 
	\sqrt{\frac{2 \gamma_\text{t} \Ki(\theta_\text{I}) \beta_\text{I}(\theta_\text{I})}  
		{\beta_\text{m}(\theta_\text{m}) + \gamma_\text{t} \beta_\text{I}(\theta_\text{I})}}\right)
	+ \frac{\gamma_\text{t} \beta_\text{I}(\theta_\text{I})}  
	{\beta_\text{m}(\theta_\text{m}) + \gamma_\text{t} \beta_\text{I}(\theta_\text{I})} \nonumber \\
	&\quad\times
	\exp\left( -
	\frac{\Km(\theta_\text{m}) \beta_\text{m}(\theta_\text{m}) 
	+ \gamma_\text{t} \Ki(\theta_\text{I}) \beta_\text{I}(\theta_\text{I})}
	{\beta_\text{m}(\theta_\text{m}) 
	+  \gamma_\text{t} \beta_\text{I}(\theta_\text{I})} \right)
	I_0\left(
	\frac{2 \beta_\text{m}(\theta_\text{m})}  {\beta_\text{m}(\theta_\text{m}) 
	+  \gamma_\text{t} \beta_\text{I}(\theta_\text{I})}
	\sqrt{\frac{\gamma_\text{t} \Km(\theta_\text{m}) \Ki(\theta_\text{I}) \beta_\text{I}(\theta_\text{I})}  {\beta_\text{m}(\theta_\text{m})}} \right)  \label{eq:Po1h}
	\end{align}
	\setcounter{eqncnt}{13}
	\centering \rule[0pt]{18cm}{0.3pt}
\end{figure*}
\setcounter{equation}{\value{eqnback}}
%
%
\section{Outage Probability Analysis}\label{sec:analytical}
In this section, 
we analyze the outage probability by considering various environments of main and interference links. 
For given the elevation angle set ${\Theta} = (\theta_\text{m},\theta_\text{I})$ and the link distance set $\mathcal{D} =  (\ell_\text{m},\ell_\text{I})$ of main and interference links,
the outage probability is defined as
\begin{align}
	\Pout(\Theta,\mathcal{D})=
	\mathbb{P}[\gamma(\theta_\text{m},\theta_\text{I}) < \gamma_\text{t}]
\end{align}
where $\gamma_\text{t}$ is the target SIR, which can be defined by $\gamma_\text{t}=2^{\frac{R_\text{t}}  {W}} - 1$ for the target rate $R_\text{t}$ and the bandwidth $W$\cite{LimYaoMan:09}.
We consider the interference limited environment, and the derived outage probabilities are given in the following theorem.
\begin{theorem}\label{trm:OPSIR}
For given ${\Theta} = (\theta_\text{m},\theta_\text{I})$ and $\mathcal{D} =  (\ell_\text{m},\ell_\text{I})$, 
the outage probability $\Pout(\Theta,\mathcal{D})$ can be presented as 
\begin{align} \label{eq:outage}
\Pout(\Theta,\mathcal{D}) 
&=
\PLoSm \PLoSi \Pout^{(\text{L,L})}(\Theta,\mathcal{D}) 
\nonumber\\
& \quad +
\PLoSm (1  -  \PLoSi) 
\Pout^{(\text{L,N})}(\Theta,\mathcal{D}) \nonumber\\
& \quad +
(1  -  \PLoSm) \PLoSi 
\Pout^{(\text{N,L})}(\Theta,\mathcal{D}) \nonumber\\
& \quad + 
(1 -  \PLoSm) (1  -  \PLoSi)
\Pout^{(\text{N,N})}(\Theta,\mathcal{D}) 
\end{align}
%
%
%
where $\Pout^{({e_\text{m},e_\text{I}})}(\Theta,\mathcal{D})$ is the outage probability 
with the environment of the main link $e_\text{m}$ and that of the interference link $e_\text{I}$. 
The environment $e_i$ can be either LoS (i.e., $e_i = \text{L}$) or NLoS (i.e., $e_i = \text{N}$), and 
$\Pout^{({e_\text{m},e_\text{I}})}(\Theta,\mathcal{D})$ for four cases of $({e_\text{m},e_\text{I}})$ are given as follows:
%
%
%
\be
\item \emph{Case} 1 ($e_\text{m}=\text{L}$ and $e_\text{I}=\text{L}$):
$\Pout\!^{(\text{L,L})}(\Theta,\mathcal{D})$ is given by \eqref{eq:Po1h}.
%
%
\setcounter{equation}{13}
\item \emph{Case} 2 ($e_\text{m}=\text{L}$ and $e_\text{I}=\text{N}$):
$\Pout\!^{(\text{L,N})}(\Theta,\mathcal{D})$ is given by
\begin{align}
	\Pout^{(\text{L,N})}(\Theta,\mathcal{D}) &= 
	\frac{\gamma_\text{t} \beta_\text{I}(\theta_\text{I})}  
	{2 \beta_\text{m}(\theta_\text{m})  +  \gamma_\text{t} \beta_\text{I}(\theta_\text{I})}
	\nonumber \\
	& \quad \times \exp\left( - 
	\frac{2 \Km(\theta_\text{m}) \beta_\text{m}(\theta_\text{m})}  
	{2 \beta_\text{m}(\theta_\text{m}) + \gamma_\text{t} \beta_\text{I}(\theta_\text{I})}\right).	\label{eq:Po2h}
\end{align}
\item \emph{Case} 3 ($e_\text{m}=\text{N}$ and $e_\text{I}=\text{L}$):
$\Pout\!^{(\text{N,L})}(\Theta,\mathcal{D})$ is given by
\begin{align}
	\Pout^{(\text{N,L})}(\Theta,\mathcal{D}) & = 
	%
	1 - \frac{\beta_\text{m}(\theta_\text{m})}  
	{2 \gamma_\text{t} \beta_\text{I}(\theta_\text{I}) + \beta_\text{m}(\theta_\text{m})} \nonumber \\
	&\quad\times 
	\exp \left( - 
	\frac{2 \gamma_\text{t} \Ki(\theta_\text{I}) \beta_\text{I}(\theta_\text{I})}  
	{2 \gamma_\text{t} \beta_\text{I}(\theta_\text{I})  +  \beta_\text{m}(\theta_\text{m})}\right).	\label{eq:Po3h}
\end{align}
\item \emph{Case} 4 ($e_\text{m}=\text{N}$ and $e_\text{I}=\text{N}$):
$\Pout\!^{(\text{N,N})}(\Theta,\mathcal{D})$ is given by
\begin{align}
	\Pout^{(\text{N,N})}(\Theta,\mathcal{D}) = 
	\frac{\gamma_\text{t} \beta_\text{I}(\theta_\text{I})}  
	{\beta_\text{m}(\theta_\text{m}) + \gamma_\text{t} \beta_\text{I}(\theta_\text{I})}.  \label{eq:Po4h}
\end{align}
\ee
\end{theorem}

\begin{IEEEproof}
The outage probability is obtained as \eqref{eq:outage} using the law of total probability.
We derive $\Pout^{({e_\text{m},e_\text{I}})}(\Theta,\mathcal{D})$ for the above four cases as follows. 
For \emph{Case} 1, $\Km(\theta_\text{m}) \neq 0$ and $\Ki(\theta_\text{I}) \neq 0$ as both main and interference links are in \ac{LoS} environments, and $\Pout^{(\text{L,L})}(\Theta,\mathcal{D})$ can be obtained
using \eqref{eq:channel fading1} as
\begin{align}
	&\Pout^{(\text{L,L})}(\Theta,\mathcal{D})
%
	=\int_{0}^{\infty} 
	\int_{0}^{\frac{\gamma_\text{t} \beta_\text{I}(\theta_\text{I}) g}  {\beta_\text{m}(\theta_\text{m})}} 
	%
	%
	f_{h_\text{m}}(h) \, dh  f_{h_\text{I}}(g)  \, dg  \nonumber \\
	&\overset{\underset{\mathrm{(a)}}{}}{=}
	%
	%
	1 - \frac{1}  {2}  \int_{0}^{\infty}  
	Q\left(\sqrt{2 \Km(\theta_\text{m})}  ,
	\sqrt{\frac{\gamma_\text{t} \beta_\text{I}(\theta_\text{I}) g}  {\beta_\text{m}(\theta_\text{m})}}\right)  \nonumber \\
	&\quad\times 
	\exp\left( - \Ki(\theta_\text{I}) 	-  \frac{g}  {2} \right)
	I_0\left(\sqrt{2 \Ki(\theta_\text{I}) g}\right) \, dg \label{eq:pout5}
\end{align}
where $Q(a,b)$ is the first-order Marcum Q-function.
In \eqref{eq:pout5}, (a) is from the \ac{CDF} of the noncentral Chi-squared distribution, and the integral term can be presented as
\begin{align}  \label{eq:marcum2}
	&\int_{0}^{\infty}  \exp\left(-c^2 x\right)  I_0\left(d \sqrt{2x}\right)  
	Q\left(e , f \sqrt{2x}\right)\,dx  \nonumber \\
	&=
	\frac{1}  {c^2}  
	\left\{
	\exp\left(\frac{d^2}  {2c^2}\right)  
	Q\left(\frac{c e}  {\sqrt{{c}^2 + {f}^2}},  
						\frac{d f}  {c \sqrt{{c}^2 + {f}^2}}\right)  \nonumber \right.\\
	&\quad \left.
	- \frac{f^2}  {c^2 + f^2}
	\exp\left(\frac{d^2 - c^2 e^2}  {2(c^2 + f^2)}\right)
	I_0\left(\frac{d e f}  {c^2 + f^2}\right)
	\right\}
\end{align}
where $c =\sqrt{0.5}$, $d =\sqrt{\Ki(\theta_\text{I})}$, $e =\sqrt{2\Km(\theta_\text{m})}$, and $f=\sqrt{\frac{\gamma_\text{t} \beta_\text{I}(\theta_\text{I})}{2 \beta_\text{m}(\theta_\text{m})}}$ from \cite[eq. (46)]{Nut:72}.  
By using \eqref{eq:marcum2} in \eqref{eq:pout5}, \scalebox{0.95}{$\Pout\!^{(\text{L,L})}(\Theta,\mathcal{D})$} is presented as \eqref{eq:Po1h}.\par

In \emph{Case} 2, $\Km(\theta_\text{m})\neq 0$ and $\Ki(\theta_\text{I})= 0$ as the interference link is in \ac{NLoS} environment, and \scalebox{0.95}{$\Pout\!^{(\text{L,N})}(\Theta,\mathcal{D})$} is obtained using \eqref{eq:channel fading1} and \eqref{eq:channel fading2} as
%
%
%
\begin{align} \label{eq:pout6}
	&\Pout^{(\text{L,N})}(\Theta,\mathcal{D}) 
%
	=\int_{0}^{\infty} 
	\int_{0}^{\frac{\gamma_\text{t} \beta_\text{I}(\theta_\text{I}) g}  {\beta_\text{m}(\theta_\text{m})}} 
	%
	%
	f_{h_\text{m}}(h) \, dh  f_{h_\text{I}}(g)  \, dg  \nonumber \\
	&=
	%
	%
	1 \hspace{-1mm} -  \hspace{-1mm}
	\int_{0}^{\infty} \hspace{-1mm}
	Q\left(\hspace{-1mm}
	\sqrt{2 \Km(\theta_\text{m})},
	\sqrt{\frac{\gamma_\text{t} \beta_\text{I}(\theta_\text{I}) g}  
	{\beta_\text{m}(\theta_\text{m})}}\right) 
	\hspace{-0.5mm}
	\exp\hspace{-0.5mm}\left(-g\right) \, dg.
\end{align}
In \eqref{eq:pout6}, the integral term can be presented as
\begin{align} \label{eq:marcum3}
	&\int_{0}^{\infty}  \exp\left(-c^2x\right)  Q\left(e , f \sqrt{2x}\right) \,dx  \nonumber \\
	&=\frac{1}  {c^2} 
	\left\{
	1 - \frac{f^2}  {c^2 + f^2}
	\exp\left(-\frac{c^2 e^2}  {2(c^2 + f^2)}\right)
	\right\}
\end{align}
where $c\hspace{-1mm}=\hspace{-1mm}1$, $e\hspace{-1mm}=\hspace{-1mm}\sqrt{2\Km(\theta_\text{m})}$, and $f\hspace{-1mm} =\hspace{-1mm}\sqrt{\frac{\gamma_\text{t} \beta_\text{I}(\theta_\text{I})}{2 \beta_\text{m}(\theta_\text{m})}}$ from \cite[eq. (40)]{Nut:72}.
By using \eqref{eq:marcum3} in \eqref{eq:pout6}, \scalebox{0.95}{$\Pout\!^{(\text{L,N})}(\Theta,\mathcal{D})$} is presented as \eqref{eq:Po2h}.\par

In \emph{Case} 3, $\Km(\theta_\text{m})= 0$ and $\Ki(\theta_\text{I})\neq 0$ as the main link is in \ac{NLoS} environment, and \scalebox{0.95}{$\Pout\!^{(\text{N,L})}(\Theta,\mathcal{D})$} is given by
\begin{align} \label{eq:pout7}
%
%
	&\Pout^{(\text{N,L})}(\Theta,\mathcal{D})
	=\int_{0}^{\infty} 
	\int_{0}^{\frac{\gamma_\text{t} \beta_\text{I}(\theta_\text{I}) g}  {\beta_\text{m}(\theta_\text{m})}} 
	%
	%
	f_{h_\text{m}}(h) \, dh  f_{h_\text{I}}(g)  \, dg  
	\nonumber \\
	&\overset{\underset{\mathrm{(a)}}{}}{=}
	%
	%
	1 -  \frac{1}  {2}  \int_{0}^{\infty}  
	\exp\left(-\frac{\gamma_\text{t} \beta_\text{I}(\theta_\text{I}) g}  {\beta_\text{m}(\theta_\text{m})}\right)  \nonumber \\
	&\quad\times
	\exp \left( - \Ki(\theta_\text{I}) - \frac{g}  {2} \right) 
	I_0	\left(\sqrt{2 \Ki(\theta_\text{I}) g}\right) \, dg  \hspace{-1mm}
\end{align}
In \eqref{eq:pout7}, (a) is from the CDF of the exponential distribution and 
the integral term can be presented as
\begin{align} \label{eq:bessel2}
	\int_{0}^{\infty}  \exp(-c^2 x)  I_0\left(d\sqrt{2x}\right)\,dx
	= 
	\frac{1}  {c^2}
	\exp\left(\frac{d^2}  {2c^2}\right)
\end{align}
where $c =\sqrt{\frac{1}{2}+\frac{\gamma_\text{t} \beta_\text{I}(\theta_\text{I})}{\beta_\text{m}(\theta_\text{m})}}$ and $d=\sqrt{\Ki(\theta_\text{I})}$ from \cite[eq. (9)]{Nut:72}.
By using \eqref{eq:bessel2} in \eqref{eq:pout7}, 
\scalebox{0.95}{$\Pout\!^{(\text{N,L})}(\Theta,\mathcal{D})$} is presented as \eqref{eq:Po3h}.

In \emph{Case} 4, $\Km(\theta_\text{m})= 0$ and $\Ki(\theta_\text{I})= 0$ as the main and the interference links are both in \ac{NLoS} environments, and \scalebox{0.95}{$\Pout\!^{(\text{N,N})}(\Theta,\mathcal{D})$} is given by
%
%
%
\begin{align}  \label{eq:pout8}
%
%
	&\Pout^{(\text{N,N})}(\Theta,\mathcal{D})
	=\int_{0}^{\infty}  
	\int_{0}^{\frac{\gamma_\text{t} \beta_\text{I}(\theta_\text{I}) g}  {\beta_\text{m}(\theta_\text{m})}} 
	%
	%
	f_{h_\text{m}}(h) \, dh  f_{h_\text{I}}(g)  \, dg
	 \nonumber \\
	&=
	%
	%
	1 - \int_{0}^{\infty}  
	\exp\left(-\frac{\gamma_\text{t} \beta_\text{I}(\theta_\text{I}) g}  {\beta_\text{m}(\theta_\text{m})} - g \right) \, dg.
\end{align}
By simple calculation in \eqref{eq:pout8}, \scalebox{0.95}{$\Pout\!^{(\text{N,N})}(\Theta,\mathcal{D})$} is obtained as \eqref{eq:Po4h}.
\end{IEEEproof}

From Theorem \ref{trm:OPSIR}, we can also obtain the outage probability as for different scenarios of \ac{UAV} communications by changing the values of $(\Theta,\mathcal{D})$.
Specifically, according to whether the main link or interference link is A2A, A2G (G2A), or G2G channel,
we can set $(\Theta,\mathcal{D})$ in \eqref{eq:outage} as the values in Table \ref{table:parameter} to obtain the outage probability in certain scenarios.

In Theorem \ref{trm:OPSIR}, we can readily know \scalebox{0.95}{$\Pout\!^{(\text{L,N})}(\Theta,\mathcal{D})$} (Case 2) 
cannot be higher than \scalebox{0.95}{$\Pout\!^{(\text{N,L})}(\Theta,\mathcal{D})$} (Case 3) as Case 2 has stronger main link and weaker interference link than Case 3. 
However, it is not clear whether the outage probability with LoS environments for both main and interference links (Case 1) can be lower or higher than that with NLoS environments for both main and interference links (Case 4). Hence, we compare \scalebox{0.95}{$\Pout\!^{(\text{L,L})}(\Theta,\mathcal{D})$} and \scalebox{0.95}{$\Pout\!^{(\text{N,N})}(\Theta,\mathcal{D})$}, and obtain the following results in Corollary 1.
%
%
%
%
%
%
\begin{table}[!t]
	\caption{$\Theta=(\theta_{\text{m}},\theta_{\text{I}})$ in outage probability \label{table:parameter}} 
	\begin{center}
		\rowcolors{2}
		{cyan!15!}{}
		\renewcommand{\arraystretch}{2}
		\begin{tabular}{ l | l  l  l }
			\hline 
			\backslashbox{Main}{Interferer} & {\hspace{0.55cm}\ac{A2A}} & {\hspace{0.1cm}\ac{A2G} (\ac{G2A})} & {\hspace{0.4cm}\ac{G2G}} \\
			\hline 
			\hspace{0.5cm} \ac{A2A}  
			& \hspace{0.3cm}  $(\frac{\pi}{2},\frac{\pi}{2})$ \hspace{0.2cm}
			& \hspace{0.2cm} $(\frac{\pi}{2},\theta_\text{I})$ & \\ 
			\hspace{0.5cm}  \ac{A2G} (\ac{G2A}) \hspace{0.1cm}
			& \hspace{0.3cm} $(\theta_\text{m},\frac{\pi}{2})$ \hspace{0.2cm}
			& \hspace{0.2cm} $(\theta_\text{m},\theta_\text{I})$ 
			& \hspace{0.2cm} $(\theta_\text{m},0)$ \\ 
			\hspace{0.5cm} \ac{G2G}  
			& & \hspace{0.2cm} $(0,\theta_\text{I})$ 
			& \hspace{0.2cm} $(0,0)$ \\ 
			\hline
		\end{tabular}\vspace{-0.4cm}
	\end{center}
\end{table}%
\begin{corollary} \label{pro:outage compare}
	According to the ratio of the average received signal power of main and interference links, i.e., $\frac{\beta_\text{m}(\theta_\text{m})}  {\beta_\text{I}(\theta_\text{I})}$, the relation between \scalebox{0.95}{$\Pout^{(\text{L,L})}(\Theta,\mathcal{D})$ and $\Pout^{(\text{N,N})}(\Theta,\mathcal{D})$} is changed as
	\begin{align}
	\left\{
	\begin{aligned}
	&
	\Pout^{(\text{L,L})}(\Theta,\mathcal{D}) > 
	\Pout^{(\text{N,N})}(\Theta,\mathcal{D}), 
	\,  \quad \text{if} \quad \frac{\beta_\text{m}(\theta_\text{m})}  {\beta_\text{I}(\theta_\text{I})}  <  v'
	\\
	&
	\Pout^{(\text{L,L})}(\Theta,\mathcal{D}) < 
	\Pout^{(\text{N,N})}(\Theta,\mathcal{D}), 
	\, 	\quad \text{if} \quad \frac{\beta_\text{m}(\theta_\text{m})}  {\beta_\text{I}(\theta_\text{I})}  >  v' \label{eq:Compare}	
	\\
	&
	\Pout^{(\text{L,L})}(\Theta,\mathcal{D})  =   
	\Pout^{(\text{N,N})}(\Theta,\mathcal{D}),
	\,  \quad \text{if} \quad \frac{\beta_\text{m}(\theta_\text{m})}  {\beta_\text{I}(\theta_\text{I})} = 0  , \infty, \, \text{or} \ v' \hspace{-10mm}
	\end{aligned}
	\right.
	\end{align}
	where $v'$ ($0< v' <\infty$) is the value of $\frac{\beta_\text{m}(\theta_\text{m})}  {\beta_\text{I}(\theta_\text{I})}$ 
	that makes \scalebox{0.95}{$\Pout\!^{(\text{L,L})}(\Theta,\mathcal{D})$}  =   
	\scalebox{0.95}{$\Pout\!^{(\text{N,N})}(\Theta,\mathcal{D})$}.
\end{corollary}
\begin{IEEEproof}
	For convenience, we introduce $v=\frac{\beta_\text{m}(\theta_\text{m})}  {\beta_\text{I}(\theta_\text{I})}$,
	and define $A(v)$ and $B(v)$ as 
	\begin{align}
	A(v)=\sqrt{\frac{2 \Km(\theta_\text{m}) v}  {v + \gamma_\text{t} }}  , \quad
	B(v)=\sqrt{\frac{2 \gamma_\text{t} \Ki(\theta_\text{I}) }  {v + \gamma_\text{t} }}.
	\label{eq:new}
	\end{align}
	%
	%
	%
	By using \eqref{eq:new}, \scalebox{0.95}{$\Pout\!^{(\text{L,L})}(\Theta,\mathcal{D})$} in \eqref{eq:Po1h} and \scalebox{0.95}{$\Pout\!^{(\text{N,N})}(\Theta,\mathcal{D})$} in \eqref{eq:Po4h} can rewrite as functions of $v$ as
	\begin{align}
	\Pout^{(\text{L,L})}(v) 
	& =  1  - Q\left(A(v), B(v)\right)
	+  \frac{\gamma_\text{t} }  {v + \gamma_\text{t} } \nonumber \\
	&\quad\times
	\exp\left( -\frac{A(v)^2  +  B(v)^2} {2}  \right) 
	I_0\left( A(v)B(v)  \right)   \nonumber \\
	\Pout\!^{(\text{N,N})}(v) &=
	\frac{\gamma_\text{t} }  {v + \gamma_\text{t}}. \label{eq:Poh_r}
	\end{align}
	%
	%
	%
	%
	%
	%
	%
	%
	%
	%
	%
	%
	%
	%
	%
	%
	%
	%
	%
	%
	%
	%
	%
	%
	%
	%
	From \eqref{eq:Poh_r}, we obtain the first derivatives of \scalebox{0.95}{$\Pout\!^{(\text{L,L})}(v)$} and \scalebox{0.95}{$\Pout\!^{(\text{N,N})}(v)$} according to $v$, respectively, as
	\begin{align}
	&\frac{\partial \Pout^{(\text{L,L})}(v)} {\partial v} 
	=  
	\left(\hspace{-0.5mm} \Pout^{(\text{N,N})}(v) - 1 \hspace{-0.5mm} \right)
	\hspace{-0.5mm}
	\exp \hspace{-0.5mm} \left(\hspace{-0.5mm} - \frac{A( v )^2  +  B( v )^2}  {2} \right) \hspace{-0.5mm}
	B(v)
	\nonumber \\
	&\times \left\{ I_1 \left( 
	A(v) B(v) \right) 
	\frac{\partial A( v )} {\partial v} 
	- I_0 \left(A(v) B(v)\right)
	\frac{\partial B(v)} {\partial v}  \right\}
	\nonumber \\
	&
	+  \Pout^{(\text{N,N})}(v)
	\exp \left( - \frac{A(v)^2 + B(v)^2}  {2} \right) 
	A(v)
	\nonumber \\
	&\times
	\left\{I_1\left(A(v)B(v)\right)
	\frac{\partial B(v)}{\partial v}
	-  I_0\left(A(v)B(v)\right)
	\frac{\partial A(v)}{\partial v} \right\} \nonumber \\
	&+ \frac{\partial \Pout^{(\text{N,N})}\hspace{-0.5mm}(v)} {\partial v}\hspace{-0.5mm}
	\exp\hspace{-1mm}\left(\hspace{-1mm}-\frac{A(v)^2 \hspace{-1mm} + 
		\hspace{-1mm} B(v)^2}  {2} \right) \hspace{-1mm}
	I_0\hspace{-1mm}\left(A(v)B(v)\right)  
	< 0   \label{eq:dif1} \\
	&\frac{\partial \Pout^{(\text{N,N})}(v)} {\partial v} = 
	- \frac{\gamma_\text{t}} {\left(v + \gamma_\text{t}\right)^2} 
	< 0 .
	\label{eq:dif2}
	\end{align}
	In \eqref{eq:dif1} and \eqref{eq:dif2}, the inequalities are obtained since
	$\exp(v) \ge 1$, $I_0(v) \ge 1$, $A(v) \ge 0$, $B(v) \ge 0$, $I_1(v) \ge 0$, $\frac{\partial A(v)}{\partial v} \ge 0$, $\frac{\partial B(v)}{\partial v} \le 0$, and $0 \le \Pout^{(\text{N,N})}(v) \le 1$.
	Hence, 
	\scalebox{0.95}{$\Pout^{(\text{L,L})}(v)$} and \scalebox{0.95}{$\Pout^{(\text{N,N})}(v)$} are
	monotonically decreasing functions of $v$.
	
	If $v = 0$, from \eqref{eq:dif2} and \eqref{eq:dif1}, we have
		\begin{align}	
		\frac{\partial \Pout^{(\text{N,N})}(0)} {\partial v} 
		<
		\frac{\partial \Pout^{(\text{L,L})}(0)} {\partial v}
		\end{align}
		since 
		$
		\frac{\partial \Pout^{(\text{N,N})}(0)} {\partial v} = 
		- \frac{1} {\gamma_\text{t}},
		$
		$
		\frac{\partial \Pout^{(\text{L,L})}(0)} {\partial v} =  
		\frac{\partial \Pout^{(\text{N,N})}(0)} {\partial v} \exp\left(-\frac{B(0)^2}  {2} \right)
		$,
		and $\Pout^{(\text{N,N})}(0) = \Pout^{(\text{L,L})}(0) = 1$.
		%
		%
		Hence, for small $\epsilon$, we have
		\begin{align}\label{eq:relation1}		
		\Pout^{(\text{N,N})}(\epsilon) < \Pout^{(\text{L,L})}(\epsilon). 
		\end{align}

	If $v$ approaches $\infty$, $B(v) \rightarrow 0 $, 
		$\lim_{v \rightarrow \infty} \Pout^{(\text{L,L})}(v) 
		= 
		\lim_{v \rightarrow \infty} \Pout^{(\text{N,N})}(v)
		= 0 
		$, 
		and from \eqref{eq:dif2} and \eqref{eq:dif1}, we have
		\begin{align} \label{eq:dif3}
		&\frac{\partial \Pout^{(\text{N,N})}(v)} {\partial v}\rightarrow
		- \frac{\gamma_\text{t}} {\left(v + \gamma_\text{t}\right)^2}, \nonumber \\
		&\frac{\partial \Pout^{(\text{L,L})}(v)} {\partial v} \rightarrow   
		\frac{\partial \Pout^{(\text{N,N})}(v)} {\partial v} \exp\left(-\frac{A(v)^2}  {2} \right).	%
		\end{align}
		From \eqref{eq:dif3}, we can see that for large $v_\text{o} \gg1$,
		$\frac{\partial \Pout^{(\text{L,L})}(v_\text{o})} {\partial v}
		>
		\frac{\partial \Pout^{(\text{N,N})}(v_\text{o})} {\partial v}$, and we have
		\begin{align} \label{eq:relation2}	
		\Pout^{(\text{L,L})}(v_\text{o}) < \Pout^{(\text{N,N})}(v_\text{o})
		\end{align}

	Therefore, from \eqref{eq:relation1}, \eqref{eq:relation2}, and the fact that 
	$\Pout^{(\text{L,L})}(v)$ and $\Pout^{(\text{N,N})}(v)$ are both monotonically decreasing functions, 
	we can know that there exists unique point $v'$ in $0<v'<\infty$ that makes 
	$\Pout^{(\text{L,L})}(v')=\Pout^{(\text{N,N})}(v')$. Therefore, we obtain \eqref{eq:Compare}.
\end{IEEEproof}

From Corollary \ref{pro:outage compare},
when the main and interference links are in the same environment,
\ac{NLoS} environment can be more preferred if the average received power of interference is much larger than that of transmitting signal (i.e., small $ \frac{\beta_\text{m}(\theta_\text{m})}  {\beta_\text{I}(\theta_\text{I})} $), but for the opposite case (i.e., large $ \frac{\beta_\text{m}(\theta_\text{m})}  {\beta_\text{I}(\theta_\text{I})} $), \ac{LoS} environment can be better 
in terms of outage probability. 
%
\section{Numerical Results}\label{sec:numerical}
In this section, we present the effects of height, parameters, and channel state on the outage probability.
Unless otherwise specified, the values of simulation parameters are $a_1\hspace{-1mm}=\hspace{-1mm}12.08$, $b_1\hspace{-1mm}=\hspace{-1mm}0.11$, $\alpha_0\hspace{-1mm}=\hspace{-1mm}3.5$, $\alpha_\frac{\pi}{2}\hspace{-1mm}=\hspace{-1mm}2$, $k_0\hspace{-1mm}=\hspace{-1mm}1$, $k_\frac{\pi}{2}\hspace{-1mm}=\hspace{-1mm}15$, and $P_\text{m}\hspace{-1mm}=\hspace{-1mm}10^{-8}W$.\par
\begin{figure}[t!]
	\begin{center}   
		{ 
			\psfrag{A}[Bl][Bl][0.59]{Outage Probability, $\Pout(\Theta,\mathcal{D})$}
			\psfrag{AAAAAAAAAAAAAAAAAAAAA}[Bl][Bl][0.59]{$\gamma_\text{t}=3,\: \ell_\text{I}=150m  \: \text{(G2A-A2A)}$}
			\psfrag{BBBBBBBBBBBBBBBBBBBBB}[Bl][Bl][0.59]{$\gamma_\text{t}=2,\: \ell_\text{I}=150m  \: \text{(G2A-A2A)}$}
			\psfrag{CCCCCCCCCCCCCCCCCCCC}[Bl][Bl][0.59]{$\gamma_\text{t}=2,\: 
			\ell_\text{I}=200m  \: \text{(G2A-A2A)}$}
			\psfrag{DDDDDDDDDDDDDDDDDDDD}[Bl][Bl][0.59]{$\text{Simulation}$}
			\psfrag{D}[Bl][Bl][0.59]{$P_\text{I}=0.1P_\text{m}$}
			\psfrag{E}[Bl][Bl][0.59]{$P_\text{I}=0.03P_\text{m}$}
			\psfrag{Y}[Bl][Bl][0.59]{$\text{Height of UAV},\: d_\text{m}^{(\text{V})} \:[m]$}
			\includegraphics[width=1.00\columnwidth]{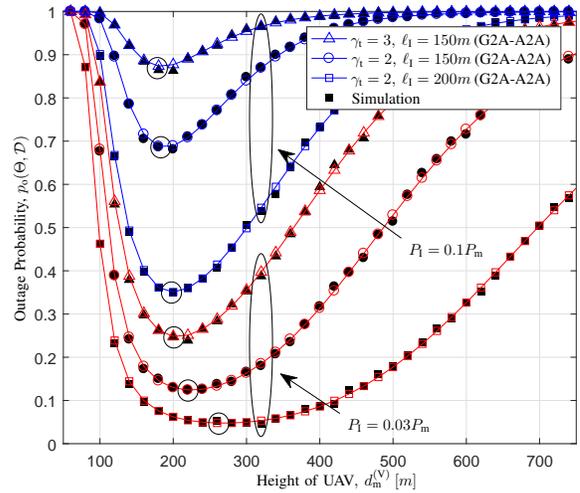}
			\vspace{-10mm}
		}
	\end{center}
	\caption{
		Outage probability $\Pout(\Theta,\mathcal{D})$ as a function of height $d_\text{m}^{(\text{V})}$ with $d_\text{m}^{(\text{H})}=100m$ for different values of $\gamma_\text{t}$, $\ell_\text{I}$, and $P_\text{I}$.
		The circle means the optimal height with the lowest outage probability.
	}
\vspace{-4mm}
	\label{fig:SIR1}
\end{figure}
%

Fig.~\ref{fig:SIR1} presents the outage probability $\Pout(\Theta,\mathcal{D})$ as a function of height $d_\text{m}^{(\text{V})}$ with $d_\text{m}^{(\text{H})}=100m$ for different values of $\gamma_\text{t}$, $\ell_\text{I}$, and $P_\text{I}$.
The main link is the \ac{G2A} channel of which the horizontal distance is fixed as 100m and the vertical distance, i.e., the height of UAV, only increases. To focus on the impact of the \ac{UAV} height on $\Pout(\Theta,\mathcal{D})$, the environment of interference link is set to be the same over different height of \ac{UAV} such as the \ac{A2A} channel with fixed link distance $\ell_\text{I}$.
In Fig.~\ref{fig:SIR1}, it is shown that the analytic results closely match with the simulation results.
%

From Fig.~\ref{fig:SIR1}, we can see that the outage probability first decreases as the height increases up to a certain value of the height, and then increases.
This is because the \ac{LoS} probability of main link increases as the height increases.
When the height of \ac{UAV} is small, as the height increases, 
the increasing probability of forming \ac{LoS} main link is more dominant than the increasing main link distance on the outage probability.
However, for large height, the \ac{LoS} probability does not change that much with the height while the link distance becomes longer, so the outage probability increases.
%
We can also see that the optimal height that minimizes $\Pout(\Theta,\mathcal{D})$ increases as the target SIR $\gamma_\text{t}$ or the power of interference link $P_\text{I}$ decreases or  the distance of interference link $\ell_\text{I}$ increases. 
From this, we can know that the optimal height increases as the impact of interference link on the communication reduces.
%
%
\begin{figure}[t!]
	\begin{center}   
		{ 
			\psfrag{A}[Bl][Bl][0.59]{$\Pout(\Theta,\mathcal{D})$}
			\psfrag{AAAAAAAAAAAAAAAAA}[Bl][Bl][0.59]{$d_\text{I}^{(\text{V})}=100m \: \text{(G2G-A2G)}$}
			\psfrag{BBBBBBBBBBBBBBBBB}[Bl][Bl][0.59]{$d_\text{I}^{(\text{V})}=50m \: \text{(G2G-A2G)}$}
			\psfrag{CCCCCCCCCCCCCCCC}[Bl][Bl][0.59]{$d_\text{I}^{(\text{V})}=100m \: \text{(A2A-G2A)}$}
			\psfrag{DDDDDDDDDDDDDDDD}[Bl][Bl][0.59]{$d_\text{I}^{(\text{V})}=50m \: \text{(A2A-G2A)}$}
			\psfrag{E}[Bl][Bl][0.59]{$\text{Horizontal distance of interference link},\: d_\text{I}^{(\text{H})} \:[m]$}
			\includegraphics[width=1\columnwidth]{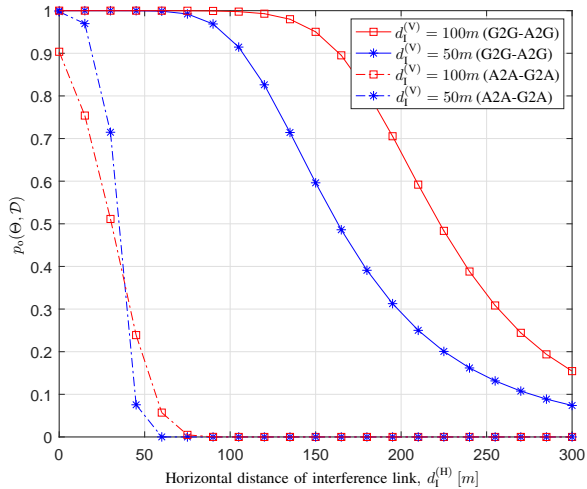}
			\vspace{-10mm}
		}
	\end{center}
	\caption{
		Outage probability $\Pout(\Theta,\mathcal{D})$ as a function of $d_\text{I}^{(\text{H})}$ with $P_\text{I}=P_\text{m}$ and $\gamma_\text{t}=2$ for different values of $d_\text{I}^{(\text{V})}$ and channel state of main link.
	}
\vspace{-4mm}
	\label{fig:SIR2}
\end{figure}
%

Fig.~\ref{fig:SIR2} presents the outage probability $\Pout(\Theta,\mathcal{D})$ as a function of $d_\text{I}^{(\text{H})}$ with $P_\text{I}=P_\text{m}$ and $\gamma_\text{t}=2$ for different values of $d_\text{I}^{(\text{V})}$ and channel state of main link. To focus on the impact of the horizontal and vertical distance of interference link, the main link is set as the \ac{A2A} or the \ac{G2G} channel with a fixed link distance 100$m$. 
The interference link is the \ac{A2G} or the \ac{G2A} channel. 
From this figure, we can see that generally, longer horizontal distance of interference link (i.e., larger $d_\text{I}^{(\text{H})}$) results in lower outage probability. On the other hand, longer vertical distance of interference link (i.e., larger $d_\text{I}^{(\text{V})}$) does not always result in lower outage probability.
Specifically, when the main link is the \ac{A2A} channel, the outage probability can be smaller with $d_\text{I}^{(\text{V})}=50m$ than with $d_\text{I}^{(\text{V})}=100m$. This is because, as $d_\text{I}^{(\text{H})}$ increases, the \ac{LoS} probability of interference link with $d_\text{I}^{(\text{V})}=50m$ decreases faster than that with $d_\text{I}^{(\text{V})}=100m$.
%
\begin{figure}[t!]
	\begin{center}   
		{ 
			\psfrag{B}[Bl][Bl][0.59]{$\Pout^{(\text{L,L})}(\Theta,\mathcal{D})$, $\Pout^{(\text{N,N})}(\Theta,\mathcal{D})$}
			\psfrag{AAAAAAAAAAAAAAAAAAAAAAAAAAAAAAAA}[Bl][Bl][0.59]{$\text{LoS main and interference links} \: \Pout^{(\text{L,L})}(\Theta,\mathcal{D})$}
			\psfrag{BBBBBBBBBBBBBBBBBBBBBBBBBBBBBBBB}[Bl][Bl][0.59]{$\text{NLoS main and interference links} \: \Pout^{(\text{N,N})}(\Theta,\mathcal{D})$}
			%
			%
			%
			\psfrag{A}[Bl][Bl][0.59]{$\frac{\beta_\text{m}(\theta_\text{m})}  {\beta_\text{I}(\theta_\text{I})}$}
			\includegraphics[width=1\columnwidth]{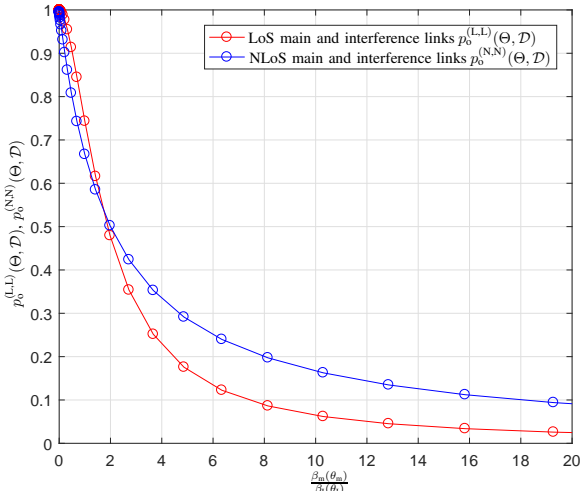}
			\vspace{-10mm}
		}
	\end{center}
	\caption{
		Outage probabilities $\Pout^{(\text{L,L})}(\Theta,\mathcal{D})$ and $\Pout^{(\text{N,N})}(\Theta,\mathcal{D})$ as a function of $\frac{\beta_\text{m}(\theta_\text{m})}  {\beta_\text{I}(\theta_\text{I})}$ with $d_\text{m}^{(\text{H})}=100m$, $d_\text{I}^{(\text{V})}=d_\text{m}^{(\text{V})}=70m$, and $\gamma_\text{t}=2$.
	}
\vspace{-4mm}
	\label{fig:SIRcompare}
\end{figure}
%

Fig.~\ref{fig:SIRcompare} presents the outage probabilities $\Pout^{(\text{L,L})}(\Theta,\mathcal{D})$ and $\Pout^{(\text{N,N})}(\Theta,\mathcal{D})$ 
as a function of $\frac{\beta_\text{m}(\theta_\text{m})}  {\beta_\text{I}(\theta_\text{I})}$ with $d_\text{m}^{(\text{H})}=100m$, $d_\text{I}^{(\text{V})}=d_\text{m}^{(\text{V})}=70m$, and $\gamma_\text{t}=2$. 
From this figure, we can confirm that both outage probabilities are monotonic decreasing functions with 
$\frac{\beta_\text{m}(\theta_\text{m})}  {\beta_\text{I}(\theta_\text{I})}$. 
In addition, there exists a cross point of those probabilities at around $\frac{\beta_\text{m}(\theta_\text{m})}  {\beta_\text{I}(\theta_\text{I})} = 1.7$. For smaller  $\frac{\beta_\text{m}(\theta_\text{m})}  {\beta_\text{I}(\theta_\text{I})}<1.7$,
$\Pout^{(\text{L,L})}(\Theta,\mathcal{D})$ is greater than $\Pout^{(\text{N,N})}(\Theta,\mathcal{D})$,
but it becomes opposite for larger $\frac{\beta_\text{m}(\theta_\text{m})}  {\beta_\text{I}(\theta_\text{I})} > 1.7$.
This verifies the results in Corollary~\ref{pro:outage compare}.

\section{Conclusion}\label{sec:conclusion}
This paper analyzes the impact of the interfering node for reliable \ac{UAV} communications. 
After characterizing the channel model affected by the elevation angle of the communication link, 
we derive the outage probability in a closed form for all possible scenarios of main and interference links. 
Furthermore, we show 
	%
the effects of the transmission power, the horizontal and vertical link distances, and  
the communication scenarios of main and interference links. 
Specifically, we show the existence of the optimal height of the \ac{UAV} for different scenarios, which increases as the power of interfering node increases or the interference link distance decreases. 
We also analytically prove that  \ac{NLoS} environment can be better than \ac{LoS} environment if the average received power of interference is much larger than that of transmitting signal.
The outcomes of our work can provide insights on the optimal deployment of \ac{UAV} in the presence of interfering node.
%
%



%

\bibliographystyle{IEEEtran}

\bibliography{StringDefinitions,IEEEabrv,mybib}

\begin{thebibliography}{10}
\providecommand{\url}[1]{#1}
\csname url@samestyle\endcsname
\providecommand{\newblock}{\relax}
\providecommand{\bibinfo}[2]{#2}
\providecommand{\BIBentrySTDinterwordspacing}{\spaceskip=0pt\relax}
\providecommand{\BIBentryALTinterwordstretchfactor}{4}
\providecommand{\BIBentryALTinterwordspacing}{\spaceskip=\fontdimen2\font plus
\BIBentryALTinterwordstretchfactor\fontdimen3\font minus
  \fontdimen4\font\relax}
\providecommand{\BIBforeignlanguage}[2]{{%
\expandafter\ifx\csname l@#1\endcsname\relax
\typeout{** WARNING: IEEEtran.bst: No hyphenation pattern has been}%
\typeout{** loaded for the language `#1'. Using the pattern for}%
\typeout{** the default language instead.}%
\else
\language=\csname l@#1\endcsname
\fi
#2}}
\providecommand{\BIBdecl}{\relax}
\BIBdecl

\bibitem{ZenZhaLim:16}
Y.~Zeng, R.~Zhang, and T.~J. Lim, ``Wireless communications with unmanned
  aerial vehicles: opportunities and challenges,'' \emph{{IEEE} Commun. Mag.},
  vol.~54, no.~5, pp. 36--42, May 2016.

\bibitem{ChoLiuLee:18}
H.~Cho, C.~Liu, J.~Lee, T.~Noh, and T.~Q.~S. Quek, ``Impact of elevated base
  stations on the ultra-dense networks,'' \emph{{IEEE} Commun. Lett.}, vol.~22,
  no.~6, pp. 1268--1271, Jun. 2018.

\bibitem{ITU:13}
P.~Series, ``Propagation data and prediction methods required for the design of
  terrestrial broadband radio access systems oprating in a frequency range from
  3 to 60 {G}hz,'' \emph{Recommendation ITU-R}, pp. 1410--1415, 2013.

\bibitem{HouKanLar:14}
A.~Al-Hourani, S.~Kandeepan, and S.~Lardner, ``Optimal {LAP} altitude for
  maximum coverage,'' \emph{{IEEE} Wireless Commun. Lett.}, vol.~3, no.~6, pp.
  569--572, Dec. 2014.

\bibitem{AzaRoChePol:18}
M.~M. Azari, F.~Rosas, K.~C. Chen, and S.~Pollin, ``Ultra reliable {UAV}
  communication using altitude and cooperation diversity,'' \emph{{IEEE} Trans.
  Commun.}, vol.~66, no.~1, pp. 330--344, Jan. 2018.

\bibitem{ZhaZhaHe:18}
S.~Zhang, H.~Zhang, Q.~He, K.~Bian, and L.~Song, ``Joint trajectory and power
  optimization for {UAV} relay networks,'' \emph{{IEEE} Commun. Lett.},
  vol.~22, no.~1, pp. 161--164, Jan. 2018.

\bibitem{CheDhi:17}
V.~V. Chetlur and H.~S. Dhillon, ``Downlink coverage analysis for a finite
  3-{D} wireless network of unmanned aerial vehicles,'' \emph{{IEEE} Trans.
  Commun.}, vol.~65, no.~10, pp. 4543--4558, Oct. 2017.

\bibitem{MozSaaBen:16}
M.~Mozaffari, W.~Saad, M.~Bennis, and M.~Debbah, ``Unmanned aerial vehicle with
  underlaid device-to-device communications: Performance and tradeoffs,''
  \emph{{IEEE} Trans. Wireless Commun.}, vol.~15, no.~6, pp. 3949--3963, Jun.
  2016.

\bibitem{WuZenZha:18}
Q.~Wu, Y.~Zeng, and R.~Zhang, ``Joint trajectory and communication design for
  multi-{UAV} enabled wireless networks,'' \emph{{IEEE} Trans. Wireless
  Commun.}, vol.~17, no.~3, pp. 2109--2121, Mar. 2018.

\bibitem{MozSaaBenDeb:17}
M.~Mozaffari, W.~Saad, M.~Bennis, and M.~Debbah, ``Mobile unmanned aerial
  vehicles ({UAV}s) for energy-efficient {I}nternet of {T}hings
  communications,'' \emph{{IEEE} Trans. Wireless Commun.}, vol.~16, no.~11, pp.
  7574--7589, Nov. 2017.

\bibitem{MorLoy:09}
V.~Mordachev and S.~Loyka, ``On node density-outage probability tradeoff in
  wireless networks,'' \emph{{IEEE} J. Sel. Areas Commun.}, vol.~27, no.~7, pp.
  1120--1131, Sep. 2009.

\bibitem{GodWie:15}
N.~Goddemeier and C.~Wietfeld, ``Investigation of air-to-air channel
  characteristics and a {UAV} specific extension to the {Rice} model,'' in
  \emph{Proc. IEEE Global Commun. Conf. Workshops. (GC Wkshps)}, San Diego, CA,
  Dec. 2015, pp. 1--5.

\bibitem{LimYaoMan:09}
W.~Limpakom, Y.~D. Yao, and H.~Man, ``Outage probability analysis of wireless
  relay and cooperative networks in {R}ician fading channels with different
  {K}-factors,'' in \emph{Proc. IEEE Veh. Technol. Conf. (VTC)}, Barcelona,
  Spain, Apr. 2009, pp. 1--5.

\bibitem{Nut:72}
A.~H. Nuttall, ``Some integrals involving the {Q}-function,'' Naval Underwater
  Systems Center (NUSC) Technical Report 4297, Tech. Rep., Apr. 1972.

\end{thebibliography}

\end{document}